\documentstyle[sprocl]{article}

\input{psfig}

\def\be{\begin{equation}}
\def\ee{\end{equation}}
\def\bea{\begin{eqnarray}}
\def\eea{\end{eqnarray}}

\begin{document}

\title{ Grand Unification and B \& L Conservation}

\author{Pran Nath}

\address{Department of Physics, Northeastern University, 
Boston,\\ MA 02115, USA\\E-mail: nath@neu.edu} 

\author{R. Arnowitt}

\address{Center for Theoretical Physics, Department of Physics,\\
Texas A \& M University, College Station, TX 77843, USA\\E-mail: 
arnowitt@phys.tamu.edu}


\maketitle\abstracts{A  review of baryon and lepton conservation
in supersymmetric grand unified theories is given. Proton stability
is discussed in the minimal SU(5) supergravity grand unification 
and in several non-minimal extensions such as the $SU(3)^3$, SO(10) and
some string based models. Effects of dark matter on proton stability
are also discussed and it is shown that the combined constraints
of dark matter and proton stability constrain the sparticle spectrum.
It is also shown that proton lifetime limits put severe constraints on
the event rates in dark matter detectors. Future prospects for the
observation of baryon and lepton number violation are also discussed.}

 \section{Introduction}
While string theory purports to unify all the fundamental 
interactions of physics including gravity, there is currently
no string model which is fully viable. Pending the discovery of
such a model, it is imperative that one look to the
bottom up approach to unification. Grand unification is such
an approach wherein one attempts to unify the electro-weak and
the strong interactions in a single framwork. In this review
we will focus on such an approach, specifically, the
approach of supersymmetric and supergravity 
grand unification\cite{grand}.
Our aim is to discuss the  phenomena of baryon
and lepton number conservation and their violation in such
theories. The grand unification approach is expected to be valid
up to the scale $M_G\sim 2\times 10^{16}$ GeV but it could be 
valid even up to the string scale of
\begin{equation}  
M_{str}\approx 5\times g_{string}\times  10^{17}~ GeV
\end{equation}
Beyond that scale one is in the domain of quantum gravity where
the full framework of string unification is needed for consistency.

 The outline of the paper is
as follows: In Sec.2 we discuss the LEP data which lends support to
the ideas of supersymmetry and gauge coupling unification. 
In Sec.3 we  give
a general discussion of the various sources of baryon number (B)
 and lepton number (L)  violation in
 SUSY theories. In Sec.4 we discuss proton decay from  B \& L violating
 dimension five operators. In Sec.5  we discuss proton decay within
 the framework of supergravity unification. In Sec.6 we discuss 
 B \& L violation in non-minimal models. These include the $(SU(3))^3$
 models, the unflipped and the flipped no-scale models, and SO(10) models.
 Effects of textures on proton lifeitme are discussed in Sec. 7
 and effects of dark matter constraints on proton lifetime are
 discussed in Sec.8. A brief discussion of proton decay in gauge
 mediated breaking of supersymmetry is given in Sec.9, and a 
 discussion of Planck effects on proton decay is given in
 Sec.10.  Exotic decay modes of the proton are discussed in
 Sec.11. In Sec.12 we give a brief discussion of the relation between
 SUSY GUTS and strings. Conclusions and prospects are given in
 Sec.13.

\section{Grand Unification and LEP Data}
While the LEP data extrapolated to high scales 
exhibits unification\cite{lang} 
in the minimal SUSY SU(5), 
 one finds that there could be a small 
 discrepancy in the predicted value of $\alpha_s$ and experiment,
 i.e., that the predicted value of $\alpha_s$ lies 
  $\sim (1-2)$ $\sigma$ higher than the world average. 
Such deviations can be corrected by Planck scale 
corrections\cite{das,ring} which
can produce $\sim (1-2)$ $\sigma$ effects on $\alpha_s$. In fact, effects
of  such a size are expected because of the proximity of the grand unification
scale to the Planck scale, and so effects of size
$O(\frac{M_G}{M_{Planck}})$, i.e., O(few\%) are quite natural\cite{hill}.
A source of such corrections in supergravity grand unification is 
from the gauge kinetic energy function $f_{\alpha \beta}$ where
\begin{equation}
[A\delta_{\alpha \beta}+\frac{c}{2M_{Planck}} d_{\alpha\beta\gamma}
\Sigma^{\gamma}]F^{\alpha}{\mu\nu}F^{\beta\mu\nu}
\end{equation}
Analyses show that values of $c\sim O(1)$ give agreement with LEP data. 
One may also use the LEP data to put a limit on the range of c .One
finds
\begin{equation}
-1\leq c\leq 3,~~~{LEP ~DATA}
\end{equation}
 It is also possible to get  1-2 $\sigma $  effects on $\alpha_s$ 
 from extensions of SU(5), such as, for example, in the missing doublet model
 with Peccei-Quinn symmetry\cite{dedes}.

\section{Sources of Baryon Number Violation in SUSY }
We dicuss now the main sources of B \& L violation in supersymmetric
theories. These consist of baryon and lepton number violation from
(i)lepto-quark exchange, (ii) dimension 4 operators, 
(iii) dimension dimension 5 operators, and (iv) higher dimensional
operators. A summary of the important operators in given in Table 1.

\begin{center} \begin{tabular}{|c|c|c|c|}
\multicolumn{4}{c}{Table~1:~ Baryon and lepton number violating operators } \\
\hline
Dim & Operator  & B \& L violation & comment\\
\hline
  dim 6   &  qqql       &    $\Delta B=1$,  $\Delta L=1$
      &           \\
\hline
   dim 9  & qqqqqq        &   $\Delta B=2$,  $\Delta L=0$      &$n-\bar n$
   ~oscillation \\
\hline
 SUSY &  &  & \\
 \hline 
dim 4     & (qqq)$_F$        & $\Delta B =1$,  $\Delta L=0$   & \\
\hline
dim 5     & (qqql)$_F$        &   $\Delta B=1$,  $\Delta L=1$      & mass scale 
$\sim 10^{16}$ GeV           \\
\hline 
dim 7 & (qqqqqq)$_F$ &  $ \Delta B=2$,  $\Delta L=0$ &$n-\bar n$ ~oscillation\\  
\hline   
\end{tabular}
\end{center}
$$ $$
The experimental possibilites for testing the baryon and lepton
number violations are (i) the 
proton decay decay experiments, (ii) experiments that look for 
double beta decay\cite{klapdor}, and (iii) proposed experiments 
for the test of 
$n-\bar n$ oscillation\cite{mohap,kamys}.
 Of these the first two are the ones that
have been most vigorously pursued. In this  review we will mostly focus
on the implications of B $\&$ L violations on proton stability.

\subsection{Proton Decay via Lepto-quark Exchange} 
As mentioned already there are various sources of proton decay in 
grand unified theories. Most grand unified theories allow for 
baryon and lepton number violation because quarks and leptons belong 
to the same multiplets and proton decay in these theories can 
proceed via lepto-quark exchange. In SU(5) models the dominant mode
via lepto-quark exchange is $p\rightarrow e^+\pi^0$ which gives a
life time for this mode of\cite{marciano} 
\begin{equation}
\tau(p\rightarrow e^+\pi^0) \approx (\frac{M_V}{3.5 \times 10^{14}  
GeV})^4 10^{31\pm 1} ~yr
\end{equation}
For the non-supersymmetric SU(5) the lifetime is too small to be 
consistent with experiment. For the supersymmetric SU(5) one estimates 
$\tau(p\rightarrow e^+\pi^0)$ to be\cite{marciano} 
\begin{equation}
 1\times10^{35\pm 1}~yr
 \end{equation}
 The current experimental limit for this decay mode is\cite{superk} 
 \begin{equation}
\tau(p\rightarrow e^+\pi^0)>2.1\times  10^{33} yr, (90\% CL)
\end{equation}
It is expected that in the future Super K will reach a 
sensitivity of\cite{totsuka} 
\begin{equation} 
\tau(p\rightarrow e^+\pi^0)>1\times  10^{34} yr, (90\% CL) 
\end{equation}
Thus the $e^+\pi^0$ mode in SUSY SU(5) may be on the edge of detection if
Super-K and Icarus reach their maximum sensitivity. 
However, as mentioned already  in suprsymmetric theories there are 
other sources of B \& L violation, such as dimension 4 and dimension 5 
operators.
 
\subsection{ p Decay  via Dimension 4 Operators }
Supersymmetric theories generically have dimension four operators which
violate baryon and lepton number. Thus, for example, in the minimal 
supersymmetric standard model (MSSM) one has in general 
dimension four operators in the superpotential of the form 
\begin{eqnarray}
W=\lambda_u Qu^cH_2 + \lambda_d Qd^cH_1+ \lambda_eLe^cH_1+
 \mu H_1H_2\nonumber\\
+(\lambda_B'u^cd^cd^c+\lambda_L'Qd^cL+\lambda_L''LLe^c)
\end{eqnarray}
Here the terms proportional to $\lambda_B'$, $\lambda_L'$, and
$\lambda_L''$  induce B \& L violation. Suppression of fast p decay 
requires 
\begin{equation}
\lambda_B'\lambda_L'<(\frac{m^2_{\tilde d}}{10^{16}GeV})^2\sim 
O(10^{-26\pm 1})
\end{equation}
In the MSSM one eliminates fast p decay via a discrete R symmetry 
$R=(-1)^{3B+L+2S}$.
However, in general R symmetry which is a global symmetry
 is not preserved by gravitational 
interactions. Thus for example, worm holes can 
generate dimension 4 operators and 
catalize p decay\cite{gilbert}.
To protect against fast p decay of the above type one must promote 
 the  global R symmetry  to a gauge symmetry, since  gauge symmetries 
 are protected against wormhole effects. Even if the local
symmetry breaks down  leaving behind  a  
residual discrete symmetry, that residual discrete symmetry will be
 sufficient  to protect against fast p decay induced by 
 worm hole effects\cite{krauss}. R parity is an interesting 
 symmetry in that it appears that it is the only $Z_2$
 symmetry which is free of anomalies with just the 
  MSSM spectrum\cite{ibanez}. Thus there is a possibility that it could 
  arise in an automatic fashion from the spontaneous breakdown
  of groups that contain a $U(1)_{B-L}$ such as
$SU(4)_C$ and  SO(10). It should be noted, however, that even if a theory 
is originally free of 
 dangerous dimension four operators, such operators can be induced 
 from higher dimensional operators via spontaneous symmetry breaking.
For example, one may have a dimension five operator in SO(10) which
contains an SU(5) singlet field $\nu^c$. The following provide 
examples where spontaneous VEV formation of an SU(5) singlet 
generates dangerous dimension four operators

\begin{equation}
\frac{1}{M_P}(u^cd^cd^c\nu^c)\rightarrow (u^cd^cd^c),
~~\frac{1}{M_P}(QLd^c\nu^c)\rightarrow (QLd^c)
\end{equation}

\section{Dimension 5 B and L Violation and p  Decay }
 In MSSM one can write many dimension 5 operators that violate B and L number
such as  $QQQL$, $u^cu^cd^ce^c$, $~QQQH_1$,$~Qu^ce^cH_1$ etc. 
These operators contribute to
p decay at the loop level. The only operators that arise  in the
minimal SU(5) GUT model and generate observable p decay are the 
first two operators in the list above\cite{wein1,acn},
 i.e.,   $QQQL$, and 
$u^cu^cd^ce^c$. \\

B and L violating dimension  five  operators occur in most 
supersymmetric theories  and
string theories and lead to p instability  in these models. In SUSY 
grand unified models proton decay via dimension five operators 
is governed by the interaction
\begin{equation}
\bar H_1J+\bar KH_1+\bar H_iM_{ij}H_j
\end{equation}
where 
$H_1, \bar H_1$ are the Higgs triplets, J and $\bar K$ are matter currents,
 and $M_{ij}$ is the Higgs triplet mass matrix. 
 The  suppression of
p  decay in these theories can come about if  
\begin{equation}
(M^{-1})_{11}=0
\end{equation}
A suppression of this type can occur by discrete symmetries, 
by non-standard embeddings, or by the presence of additional Higgs 
triplets\cite{testing,gomez}. However, in most SUSY/string models 
(except for  the flipped $SU(5)\times U(1)$ models) one does not have a
a natural suppression, and a suppression requires a  doublet-triplet 
splitting in the Higgs multiplets. Many  mechanisms for doublet-triplet 
splittings have been discussed in the literature, such as
 (i) the sliding singlet mechanism\cite{barr} which works 
for SU(n) for n$\geq 6$,
(ii) the  missing  partner mechanism\cite{grin}, (iii) the mechanism of
 VEV alignment\cite{dimo}, (iv) the mechanism where the Higgs doublets are 
 pseudo-Goldstones\cite{gift}, 
 and (v) the mechanism with more than one adjoint Higgs.
 
 In the following we will discuss  p decay via dimension 5 operators 
 in several SUSY  GUTS: SU(5), (SU(3))$^3$, SO(10), the flipped and the
 unflipped no-scale models, with the most emphasis on the simplest SUSY GUTs, 
 i.e., the minimal SU(5) model. The p decay  in the minimal SU(5) model
 is governed by 

\begin{equation}
W_Y=-\frac{1}{8}f_{1ij}\epsilon_{uvwxy}H_1^uM_i^{vw}M_j^{xy}+
f_{2ij}\bar H_{2u}\bar M_{iv} M_j^{uv}
\end{equation}
where $M_{ix}$ and $M_i^{xy}$ (i=1,2,3) are the 
$\bar 5$ and 10 of SU(5) which contain the three 
generations of quarks and leptons, and
$H_1, H_2$ are the $\bar 5$,5 of Higgs.  
After the breakdown of the GUT symmetry there is a splittiing of the 
Higgs multiplets where the Higgs triplets become superheavy and 
the Higgs doublets remain light by one of the mechanisms listed 
earlier. One can now integrate on the Higgs triplet field and
obtain an effective interaction at low energy given below 
\begin{eqnarray}
\it LLLL= \frac{1}{M} \epsilon_{abc}(Pf_1^uV)_{ij}(f_2^d)_{kl}
( \tilde u_{Lbi}\tilde d_{Lcj}(\bar e^c_{Lk}(Vu_L)_{al}-
\nu^c_kd_{Lal})+...)+H.c.\nonumber\\
 RRRR= -\frac{1}{M} \epsilon_{abc}(V^{\dagger} f^u)_{ij}(PVf^d)_{kl}
(\bar e^c_{Ri}u_{Raj}\tilde u_{Rck}\tilde d_{Rbl}+...)+H.c.
\end{eqnarray}
where  V is the CKM matrix and $f_i$ are the Yukawa couplings which are
 related to the quark masses by 
\begin{equation} 
m_i^u=f_i^u (sin2\theta_W/e)M_Z sin\beta,~~
m_i^d=f_i^d (sin2\theta_W/e)M_Z sin\beta 
\end{equation}
and $P_i$ are generational phases 
~~\\
\begin{equation}
P_i=(e^{i\gamma_i}), ~\sum_i \gamma_i=0; ~i=1,2,3
\end{equation}
 Both LLLL and RRRR interactions must be taken into account in a full 
 analysis and their relative strength depends on the part of the parameter
 space where their effects are computed\cite{acn,goto}. 

The operators  of Eq.(14) are dimension five operators which 
must be dressed via the  exchange of  
gluinos, charginos and neutralinos. The dressings give 
rise to  dimension six operators. These dimension six operators are
then used in the computation of proton decay.
In the dressings one takes into account the L-R mixings,
where for the up squark mass matrix one has
\begin{equation}
\left({{{m_{{Ru}}^{2}}\atop{m_u(A_{u}m_0-\mu
ctn\beta)}}{{m_u(A_um_0-\mu ctn\beta)}\atop{m_{Lu}^{2}}}}\right)
\end{equation}
The mass diagonal states are given by 
\begin{eqnarray}
\tilde u_{R}=cos\delta_{u}\tilde u_{1}+sin\delta_{u}\tilde u_{2},
~\tilde u_{L}=-sin\delta_{u}\tilde u_{1}+
cos\delta_{u}\tilde u_{2}\nonumber\\
sin2\delta_{u}=-2 m_{u}(A_{u}m_0-\mu ctn\beta)/(\tilde m_{u1}^2-\tilde
m_{u2}^2)
\end{eqnarray}
and similarly for the down squarks and the leptons. In supergravity grand 
unification the values of the trilinear couplings $A_u, A_d, A_e$ 
are all related to the single parameter $A_0$. After 
dressing of the dimension 5 by the  gluino, the chargino 
and the neutralino 
exchange diagrams one finds baryon and lepton number violating  
dimension six operators with chiral structures  LLLL, LLRR, RRLL and RRRR
which enter in the  proton decay analysis.

\subsection{Effective Lagrangian Approach }
The B and L violating interaction one gets from the fundamental SU(5)
Lagrangian is in terms of quarks and leptons, while the p decays 
involve physical mesons and baryons. There are various techniques for
bridging the gap, i.e., in going from the fundamental to the 
phenomenological  interactions, such as the Bag model, lattice QCD, 
and effective Lagrangians. Currently  the most efficient of these 
approaches is that of effecive lagrangians\cite{chadha}. 
Here the basic technique consists
in finding the effective  interactions in terms of mesons 
and baryons with the same
chiral structures as  LLLL, LLRR, etc. that appear in the 
fundamental Lagrangian. This effort is facilitated by 
utilizing the transformation properties of these terms under
$SU(3)_L\times SU(3)_R$ as shown below: 
\noindent
$$ Chiral ~structure~~~~~~~~~ SU(3)_L \times SU(3)_R ~rep$$
$$LLLL~~~~~~~~~~~~~~~~~~~~~~~~~~~~~~~(8,1)$$ 
$$LLRR~~~~~~~~~~~~~~~~~~~~~~~~~~~~~~~(3^*,3)$$
$$RRLL~~~~~~~~~~~~~~~~~~~~~~~~~~~~~~~(3,3^*)$$
$$RRRR~~~~~~~~~~~~~~~~~~~~~~~~~~~~~~~(1,8)$$
In the effective lagrangian approach one finds combinations of 
mesonic and baryonic fields with the same chiral transformation
properties as the dimension six  B\& L violating quark fields.
For this purpose one defines first  a pseudo-goldstone mass matrix

\begin{equation}
M=\pmatrix{\offinterlineskip
 \frac{\pi^0}{\sqrt{2}}+\frac{\eta}{\sqrt{6}}&\pi^+&K^+\cr
\pi^-&     -\frac{\pi^0+}{\sqrt{2}}+\frac{\eta}{\sqrt{6}}   &K^0\cr
K^-& \bar K^0 & -\sqrt{\frac{2}{3}} \eta\cr}
\end{equation}
Similarly, one defines a baryon mass matrix so that 
\begin{equation}
B=\pmatrix{\offinterlineskip
 \frac{\Sigma^0}{\sqrt{2}}+\frac{\Lambda}{\sqrt{6}}&\Sigma^+&p\cr
\Sigma^-&     -\frac{\Sigma^0+}{\sqrt{2}}+\frac{\Lambda}{\sqrt{6}}   &n\cr
\Xi^-& \Xi^0 & -\sqrt{\frac{2}{3}} \Lambda\cr}
\end{equation}

\noindent
Defining 
\begin{equation}
\xi=e^{i\frac{M}{f}}
\end{equation}
one can find combinations of mesons and baryon mass matrices with
the  following transformations

$$ meson-baryon ~structure~~~~~~~~~ SU(3)_L \times SU(3)_R ~rep$$
$$\xi B \xi^{\dagger}~~~~~~~~~~~~~~~~~~~~~~~~~~~~~~~~~~(8,1)$$ 
$$\xi^{\dagger} B \xi^{\dagger}~~~~~~~~~~~~~~~~~~~~~~~~~~~~~~~~~~(3^*,3)$$ 
$$\xi B \xi~~~~~~~~~~~~~~~~~~~~~~~~~~~~~~~~~~~(3,3^*)$$ 
$$\xi^{\dagger} B \xi~~~~~~~~~~~~~~~~~~~~~~~~~~~~~~~~~~(1,8)$$ 
Using the above technique one can write an effective Lagrangian 
with the same chiral transformation properties as the fundamental
Lagriangian in terms of quarks. 
Currently the effective lagrangian approach is the most reliable 
approach to the computation of proton decay amplitudes.

In the minimal SU(5) model the 
 dominant decay modes of the proton involve pseudo-scalar bosons and
anti-leptons, i.e., 
\begin{eqnarray}
\bar\nu_iK^+,\bar\nu_i\pi^+ ; i=e,\mu,\tau\nonumber\\
e^+K^0,\mu^+K^0,e^+\pi^0,\mu^+\pi^0,
e^+\eta,\mu^+\eta
\end{eqnarray}
The  relative strength of these decay modes depends on various
factors, such as quark masses, CKM  factors,
and third generation effects in the loop diagrams etc. denoted by $y^{tk}_1 $
below.  The various decay modes and some of the factors that control
these decays  modes are summarized in Table 1.
$$  $$
\begin{center} \begin{tabular}{|c|c|c|}
\multicolumn{3}{c}{Table~1:~lepton + pseudoscalar decay modes of the proton } \\
\hline
Mode & quark factors  & CKM factors \\
\hline
$\bar \nu_eK$ &$m_d m_c$  &$V_{11}^{\dagger}V_{21}V_{22} $\\
\hline
$\bar \nu_\mu K $ &$m_s m_c$  &$V_{21}^{\dagger}V_{21}V_{22} $\\
\hline
$\bar \nu_\tau K $ &$m_b m_c$  &$V_{31}^{\dagger}V_{21}V_{22} $\\
\hline
$\bar \nu_e \pi,\bar \nu_e \eta $ &$m_d m_c$  &$V_{11}^{\dagger}V_{21}^2 $\\
\hline
$\bar \nu_\mu \pi,\bar \nu_\mu \eta$ &$m_s m_c$  &$V_{21}^{\dagger}V_{21}^2 $\\
\hline
$\bar \nu_\tau \pi,\bar \nu_\tau\eta $ &$m_b m_c$  &$V_{31}^{\dagger}V_{21}^2 $\\
\hline
$eK $ &$m_d m_u$  &$V_{11}^{\dagger}V_{12} $\\
\hline
$\mu K $ &$m_s m_u$  & \\
\hline
$e\pi, e\eta $ & $m_d m_u$ &   \\
\hline
$\mu \pi,\mu \eta $ &$m_s m_u$  & $V_{11}^{\dagger}V_{21}^{\dagger}$\\
\hline
\end{tabular} 
\end{center}
$$ $$ 
The  order of magnitude estimates can be gotten using    
\begin{equation}
m_uV_{11}:m_cV_{21}:m_t V_{31}\approx 1:50:500
\end{equation}
Typically the most dominant mode is $\bar\nu K$.
It is governed by the interaction 

\begin{eqnarray}
L_6(N\rightarrow \bar\nu_i K) =((\alpha_2)^2 (2MM_W^2sin2\beta)^{-1}
P_2m_cm_i^d V_{i1}^{\dagger}V_{21} V_{22})\nonumber\\
  (F(\tilde c,\tilde d_i, \tilde W)
+F(\tilde c\tilde d_i, \tilde W)[(1+y_i^{tK}+(y_{\tilde g}
+y_{\tilde Z})\delta_{i2}+ \Delta_i^K)\alpha_i^L\nonumber\\
+(1+y_i^{tK}-(y_{\tilde g}
y_{\tilde Z})\delta_{i2}+\Delta_i^K)\beta_i^L
+(y_1(R)\alpha_3^R+y_2^{(R)}\beta_3^R)\delta_{i3}])
\end{eqnarray}
where
\begin{eqnarray}
\alpha_i^L=\epsilon_{abc}(d_{aL}\gamma^0u_{bL})(s_{cL}
\gamma^0\nu_{iL})\nonumber\\
\alpha_i^R=\alpha_i^L(d_L,u_L\rightarrow d_R,u_R)\nonumber\\
 ~\beta_i^{L,R}=
\alpha_i^{L,R}(d\leftarrow\rightarrow s)
\end{eqnarray}
In the above 
$y_i^{tk}$ is the third  generation  contribution 
\begin{equation}
y_i^{tK}=\frac{P_2}{P_3}(\frac{m_tV_{31}V_{32}}{m_c V_{21}V_{22}})
(\frac{F(\tilde t,\tilde d_i,\tilde W)+F(\tilde t,\tilde e_i,\tilde W)}
{F(\tilde c,\tilde d_i,\tilde W)+F(\tilde c,\tilde e_i,\tilde W)}) 
\end{equation}
where F are dressing loop integrals and 
 $y_{\tilde g}$ are corrections from the gluino exchange,
 and $y_{\tilde Z}$ are corrections from neutralino
exchange. The second and the third generation squark loop
contributions can interfere both constructively and destructively.
The p lifetime is enhanced when there is destructive interference. 
Further, there are situations when  the $\bar \nu \pi^+$ mode 
can be significantly enhanced so that it becomes comparable to
the $\bar \nu K^+$ mode\cite{acn}.

\subsection{The $\bar\nu K$ Mode}
We discuss now the details of the $\bar\nu K$ mode which as 
already pointed out is most
often the most dominant mode in the nucleon decay in the minimal
SU(5) model. The decay width of the $p\rightarrow \bar \nu_i K$
mode is given by 
\begin{equation}
\Gamma(p\rightarrow \bar\nu_iK^+)=(\frac {\beta_p}{M_{H_3}})^2|A|^2
|B_i|C
\end{equation}
Here  the factors A and $B_i$ are given by  
\begin{equation}
A=\frac{\alpha_2^2}{2M_W^2}m_s m_c V_{21}^{\dagger} V_{21}A_L A_S
\end{equation}

\begin{equation}
B_i= \frac{1}{sin2\beta}\frac{m_i^d V_{i1}^{\dagger}}{m_sV_{21}^{\dagger}} 
[P_2 B_{2i}+\frac{m_tV_{31}V_{32}}{m_cV_{21}V_{22}} P_3B_{3i}]
\end{equation}

\begin{equation}
B_{ji}=F(\tilde u_i,\tilde d_j,\tilde W)+(\tilde d_j\rightarrow 
\tilde e_j)
\end{equation}

\begin{eqnarray}
F(\tilde u_i,\tilde d_j,\tilde W)=[E cos\gamma_-sin\gamma_+\tilde f(\tilde
u_i,\tilde d_j, \tilde W_1)
+cos\gamma_+sin\gamma_-\tilde f(\tilde
u_i,\tilde d_j, \tilde W_1)]\nonumber\\
-\frac{1}{2} \frac{\delta_{i3}m_i^u sin2\delta_{ui}}{\sqrt 2 M_W sin\beta}
[E sin\gamma_-sin\gamma_+\tilde f(\tilde
u_{i1},\tilde d_j, \tilde W_1)
-cos\gamma_-cos\gamma_+\tilde f(\tilde
u_{i1},\tilde d_j, \tilde W_2)\nonumber\\
 - (\tilde u_{i1}\rightarrow \tilde u_{i2})]
\end{eqnarray}
where $\tilde f$ is given by 
\begin{equation}
\tilde f(\tilde u_i,\tilde d_j, \tilde W_k)=sin^2\delta_{ui}
\tilde f(\tilde u_{i1},\tilde d_j, \tilde W_k)
+cos^2\delta_{ui}
\tilde f(\tilde u_{i2},\tilde d_j, \tilde W_k) 
\end{equation}
and where 
\begin{equation}
f(a,b,c)=\frac{m_c}{m_b^2-m_c^2}[\frac{m_b^2}{m_a^2-m_b^2}ln(\frac{m_a^2}
{m_b^2})-(m_a\rightarrow m_c)] 
\end{equation}
In Eq.(30) $ \gamma_{\pm}=\beta_+\pm\beta_- $ and 
\begin{eqnarray}
sin2\beta_{\pm}=\frac{(\mu\pm \tilde m_2)}{[4\nu_{\pm}^2
+(\mu\pm \tilde m_2)^2]^{1/2}}\nonumber\\
\sqrt 2 \nu_{\pm}=M_W(sin\beta\pm cos\beta)\nonumber\\
sin2\delta_{u3}=-\frac{-2(A_t+\mu ctn\beta)m_t}{m_{\tilde t_1}^2-
m_{\tilde t_2}^2}\nonumber\\
E=1~, sin2\beta>\mu\tilde m_2/M_W^2\nonumber\\
~~~=-1,sin2\beta<\mu\tilde m_2/M_W^2 
\end{eqnarray}
C  in Eq.(27) is  a current algebra factor and is given by 
\begin{equation}  
C=\frac{m_N}{32\pi f_{\pi}^2} [(1+\frac {m_N(D+F)}{m_B})
(1-\frac{m_K^2}{m_N^2})]^2
\end{equation}
where  $f_{\pi}, D,F, ..$ etc are  the chiral Lagrangian factors
with the  the numerical values:
$ f_{\pi}=139$~MeV,D=0.76,F=0.48,$m_N$=938 ~MeV, ~$m_K$=495 ~MeV, 
and  ~$m_B$=1154.Finally,  in Eq.(27) $\beta_p$ is defined by
\begin{equation}
\beta_p U_L^{\gamma}=\epsilon_{abc}\epsilon_{\alpha \beta} <0|d_{aL}^{\alpha}
u_{bL}^{\beta}u_{cL}^{\gamma}|p>
\end{equation}
where the lattice gauge analysis gives\cite{gavela} 
\begin{equation}
\beta_p=(5.6\pm 0.5)\times  10^{-3} GeV^3
\end{equation}

\subsection{Vector Meson Decays Modes of the Proton }
 In addition to the nucleon decay modes involving pseudo-scalar
bosons and anti-leptons, one also has in general decay modes
involving vector bosons and anti-leptons. The source of these
modes are the same baryon number violating dimension six quark operators 
that  give rise to the decay modes that give rise to 
pseudoscalar and ant--lepton modes. 
The vector decay modes of the proton are
\begin{eqnarray}
\bar\nu_iK^*,\bar\nu_i\rho,\bar\nu_i\omega ; i=e,\mu,\tau\nonumber\\
e K^*,\mu K^*,e\rho,\mu\rho,e\omega,\mu\omega\nonumber\\
\end{eqnarray}
However, the vector meson decay modes have generally smaller branching 
ratios than  the corresponding pseudo-scalar decay modes\cite{yuan}.

\section{Supergravity Analysis}
We shall  work here in the framework of supergravity models where
supersymmetry is  broken in the hidden  sector by
a superhiggs phenomenon and the breaking 
communicated gravitationally to the 
physical sector\cite{chams,applied}. 
The simplest case corresponds to
when the superhiggs coupling are  generation blind. Here after 
breaking of supersymmetry and of the gauge group and after  
integrating out the  superhiggs fields  and the heavy fields of the theory,
one finds that the supersymmetry breaking 
potential below the GUT scale is given by 
\begin{equation}
V_{SB}=m_0^2 z_az_a^{\dagger}+(A_0W^{(3)}+B_0W^{(2)} +h.c.)
\end{equation}
and in addition one has a universal gaugino mass term
\begin{equation}
\it L^{\lambda}_{mass}=-m_{1/2}\bar \lambda^{\alpha}\lambda^{\alpha}
\end{equation}
In the above, $W^{(3)}$ is the trilinear part of the 
superpotential and $W^{(2)}$ is the bilinear part which under the constaint of R parity 
invariance is given by $W^{(2)}=\mu_0 H_1H_2$.  Thus we find that the
supersymmetry breaking sector contains just four parameters.

	 Supergravity unification possesses the remarkable feature that 
	  the electro-weak symmetry  breaking can be manufactured by radiative 
	  effects. The radiative electro-weak symmetry breaking is 
	  governed by the 
	  following Higgs potential 
	  \begin{eqnarray}
V_H=m_1^2(t)|H_1|^2+m_2^2(t)|H_2|^2-m_3^2(t)(H_1 H_2 +h.c.)\nonumber\\
~~~~+\frac{1}{8} (g^2+g_y^2)(|H_1|^2-|H_2|^2)^2 +\Delta V_1
\end{eqnarray}
where $\Delta V_1$ is the one loop correction to the Higgs
 potential and the parameters
$m_i^2(t)$  satisfy  
\begin{equation}
m_i^2(0)=m^2_0+\mu^2_0; i=1,2;~~m_3^2(0)=-B_0\mu_0
\end{equation}
while the gauge coupling constants satisfy the GUT relation
$\alpha_3(0)$=  $\alpha_2(0)=\alpha_G=(5/3)\alpha_Y(0)$.  
The breaking of the electroweak symmetry is accomplished by  
a satisfaction of the constraints ($\tan\beta=<H_2>/<H_1>)$:
\begin{eqnarray}
\frac{1}{2}M_Z^2=(\mu_1^2-\mu_2^2 tan^2\beta)/(tan^2\beta-1)\nonumber\\
~sin2\beta=(2m^2_3)/(\mu_1^2+\mu_2^2)\nonumber\\
\mu_i^2\equiv m_i^2+\Sigma_i
\end{eqnarray}
where $\Sigma_i$ is the one loop correction from $\Delta V_1$.
 On using the radiative
breaking constraint the low energy SUSY parameters can be 
chosen to be 

\begin{equation}
 m_0, m_{1/2}, A_0, tan\beta, sign{\mu}
\end{equation}

Supergravity unification exhibits the phenomenon of scaling over most of the 
parameter space of the model. This arises  because over most of the parameter space 
of the model one has $\mu^2>>M_Z^2$ which gives\cite{scaling}
\begin{eqnarray}
m_{\tilde W_1}\sim \frac{1}{3} m_{\tilde g}~( \mu<0); ~~
m_{\tilde W_1}\sim \frac{1}{4} m_{\tilde g}~(\mu>0)\nonumber\\
2 m_{\tilde Z_1}\sim m_{\tilde W_1}\sim m_{\tilde Z_2};~~
m_{\tilde Z_3}\sim m_{\tilde Z_4}\sim m_{\tilde W_2} >> 
m_{\tilde Z_1}
\end{eqnarray}
In addition one also has $m_{H^0}\sim m_A\sim m_{H^{\pm}}>>m_h$. 
Corrections to scaling  are  O($M_Z$/$\mu$). While these corrections are small over
most of the parameter space of the model, they can become significant in the region of
small $\mu$.  

\subsection{Non-universalities of Soft SUSY Breaking Parameters}
Universalities of the soft SUSY breaking parameters at the GUT scale
are a consequence of the assumption of a generation independent Kahler
potential in the supergravity analysis. 
However, more generally one can have generational 
dependence\cite{soni} and 
 string based analyses indeed show such a dependence.
 From the phenomenological view point,
these generational dependences cannot be arbitrary but
are severly constrained by flavor changing neutral currents (FCNC).
However, it is possible to introduce significant amounts of
 non-universalities in the Higgs sector and in the third generation 
 sector without upsetting the FCNC constraints\cite{matallio,nonuni,datta}.
 Thus, for example, Higgs sector non-universalities can be 
 parametrized at the GUT scale so that 
 \begin{equation}
 m_{H_1}^2=m_0^2(1+\delta_1),  
 m_{H_2}^2=m_0^2(1+\delta_2)  
\end{equation}
where a reasonable range for $\delta_i$ is $|\delta_i|\leq 1$.
 Similarly one can parametrize the non-universalities in the 
 third generation sector.

\begin{figure}
\vskip -1.5cm
\psfig{figure=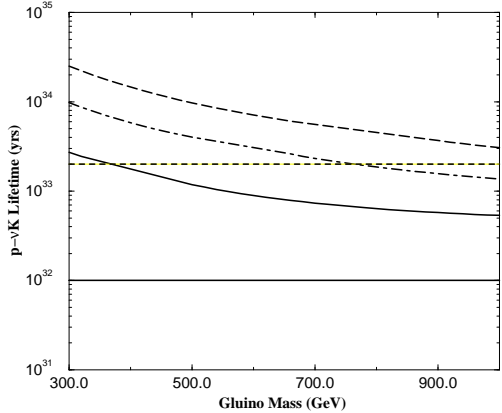,height=2.5in}
\caption{ Plots of the maximum $\tau(p\rightarrow \bar \nu K)$ lifetime
	in the minimal SU(5) 
	supergravity model with universal soft breaking as a function
	of the gluino mass. The analysis is for the naturalness limits 
	on $m_0$ of 
	1 TeV (solid), 1.5 TeV (dashed-dot) and 2 TeV (dashed). The 
	 horizontal solid line is the lower limit
	from Kamionkande and the horizontal dashed line is the lower limit 
	expected from Super K. (Taken from ref.[35]).
\label{fig:radis1} }
\end{figure}

\subsection{p Decay in mSUGRA}
We give here a numerical analysis of the maximum p lifetime for the
minimal supergravity model. The upper bounds on the proton lifetime
for different values of $m_0$ as a function of the gluino mass 
are given in Fig.1. These upper limits 
 are gotten by integrating over the allowed range
of the parameter space, i.e., $\tan\beta$ and $A_0$ for fixed
values of $m_0$ and $m_{\tilde g}$,  consistent with radiative breaking
of the electro-weak symmetry. The analysis is carried out under the 
constraint on the Higgs triplet mass such that $M_{H_3}\leq 10 M_G$.
The theoretical upper bounds may be compared with the current experimental
lower bound on the $p\rightarrow\bar \nu K^+$ mode\cite{superk} of  
 \begin{equation}
 \tau(p\rightarrow \bar\nu K)>5.5\times 10^{32}~yr
 \end{equation}
and the lower limits that the super Kamiokande (SuperK)
 hopes to achieve in
the future for this mode which is\cite{totsuka}
\begin{equation}
 \tau(p\rightarrow \bar\nu K)>2\times 10^{33}~yr
 \end{equation}
The analysis shows that the  current lower limit on 
the $\bar\nu K^+$ mode does not exhaust the parameter space for the
naturalness limit of $m_0\leq 1$ TeV. However, most of  the 
parameter space for this naturalness limit will be exhausted
when SuperK reaches its expected sensitivity of $2\times 10^{33}$ yr.

\section{B and L Violation in Non-minimal Models }
In this section we will discuss B and L violation in various 
non-minimal models such as the $(SU(3))^3$ models, unflipped 
and flipped no-scale models, and SO(10) models. 

\subsection{$(SU(3))^3$ Models}
$(SU(3))^3$  models arise naturally in many Calabi-Yau string model
constructions which below the compactification scale have the
gauge structure $E_6\times N=1~~supergravity$. 
After Wilson line  breaking, the $E(6)$ can break to $(SU(3))^3$
and one can make contact with low energy physics. 
The most studied models of this type are the three generation models 
such as $CP^3\times CP^3/Z_3$, and $CP^2\times CP^3/Z_3\times Z_3'$. In 
 models of this type the massless sector of the theory falls into 
 $27 + \bar {27}$ of $E_6$. The 27-plet  decomposes under 
 $SU(3)_C\times SU(3)_L\times SU(3)_R$ so that (see, e.g, 
 refs.\cite{greene,gepner,calabi}), 
   
 \begin{equation}
27=L(1,3,\bar 3) + Q(3,\bar 3, 1) + Q^c(\bar 3, 1, 3)
\end{equation}
and one has a similar decomposition of $\bar {27}$ into 
$\bar L$+ $\bar Q$ + $\bar Q^c$.  The particle content of 
L, Q, Q$^c$ is as follows
\begin{equation}
L=( \it l, H, H', e^c, \nu^c, N);~~Q=(q, D);~~Q^c=(q^c, D^c)
\end{equation}
where $\it l, q,..$  etc are the  $SU(2)_L$ doublets and $D$ and $D^c$ are
 the $SU(2)_L$  singlet quarks. The coupling structure of the theory is
 given by 
 \begin{eqnarray}  
(27)^3=\lambda_1 d u D +\lambda_2 u^c d^c D^c
  +\lambda_3[-HH'N-H\nu^c\it l + H'e^c\it l]\nonumber\\
~~~~~+\lambda_4[D N D^c -De^c  u^c +D^a \nu^c d^c
+q\it lD^c-qHu^c-qH'd^c]
\end{eqnarray}
There are several sources of proton instabiltiy in the these.
We list the two dominant sources among these:
(a) From B and L violating interactions which contain 
 a D or a $D^c$. They lead to nucleon decay from the 
 B and L violating dimension 5 operators after D and $D^c$ terms are
 eliminated; 
 (b) $D-d$ mixing arising from  N and $\nu^c$ VEV growth.
 This interaction provides the other dominant source of p decay 
 in this theory. The superpotential  with   
 the D-d mixing after N and $\nu^c$ VEV growth has the form 
\begin{equation}
 W_3^{D-d}=[M_1 D D^c +M_2 D d^c + M_3 d d^c]
 \end{equation}
 Diagonalization of the mass matrix leads to interactions which 
 lead to proton decay\cite{pcalabi}. \\

Superstring models of Calabi-Yau type with $(SU(3))^3$ gauge 
group can 
generate  neutrinoless double beta decay 
 $(\beta\beta)_{0\nu}$ after R parity violation. Thus 
as discussed $\nu^c$ VEV growth violates lepton number
 and R parity and one
can obtain an effective Lagrangian of the type 
\begin{equation}
L_{eff}=g \bar\chi_1^0 e_L\bar d_a u^a_L
\end{equation}
 Since $\bar\chi_1^0$ is Majorana, the interaction allows
a neutrinoless double beta decay. The predictions give\cite{tufts} 
\begin{equation}
g^2/m_{\chi_1^0}<10^{-15} GeV^{-5}
\end{equation}
while the current experimental limits correspond to 
\begin{equation}
g^2/m_{\chi_1^0}<10^{-13} GeV^{-5}
\end{equation}
Thus the predicted level of double beta decay in Calabi-Yau 
models lies below the current level of sensitivity of 
experiment by about two orders of magnitude.

\subsection{P Decay in No Scale Models }
An important sub-class of supergravity models are the so-called
no scale models. In such models the F term contribution to 
the scalar potential\cite{applied} of the
models, i.e.,  
\begin{equation} 
V=e^{-G} [G^{-1,B}_{,A}G_{,B}G^{,A} -3]+\frac{1}{2}
ReF_{\alpha\beta}^{-1}D^{\alpha}D^{\beta}
\end{equation}
vanishes because one in on an Einstein manifold. 
In a class of no-scale models one has $m_0=0=A_0$ and 
SUSY breaking is driven by the universal gaugino mass
term $m_{1/2}$. In this class of models one typically 
finds  $m_{\tilde q}< m_{\tilde g}$.
Unfortunately unflipped no-scale models of the above type violate 
experimental p lifetime limits\cite{noscale}
 in the interesting part of the parameter 
space since proton decay is governed roughly by the ratio
 $m_{\tilde g}/m_{\tilde q}^2$, and the  constraint 
$m_{\tilde q}< m_{\tilde g}$ tends to destalilize the
proton.

\subsection{Proton Decay in the Flipped SU(5)  Models}
In the flipped SU(5) model\cite{flipped}
 some particles assigned to the $\bar 5$-plets ($\bar 5_i;i=1,2,3)$
  and the 10-plets ($T_i; i=1,2,3)$) 
are flipped, i.e., 
\begin{equation}
u^c\leftarrow\rightarrow d^c,~~~e^c\leftarrow\rightarrow \nu^c
\end{equation}
Thus one has three generations of matter of the type
\begin{equation}
\bar F_i=(u^c_i,L_i=(e_i,\nu_{e_i}));~~T_i=(q_i=(u_i,d_i), d^c_i,\nu^c_i);
~~\it e^c_i;~~i=1-3
\end{equation}
The Higgs sector consists of the usual one pair of $5+\bar 5$ of Higgs
$h,\bar h$, i.e., 
\begin{equation}
 h=(H,D),~~\bar h=(\bar H, \bar D)
\end{equation}
 and two pairs of $10,\bar{10}$ of Higgs $H_i, \bar H_i$(i=1,2)
so that
\begin{equation}
H_i=(q_{H_i}, d_{H_i}^c,\nu_{H_i}^c),~~~
\bar H_i=(q_{\bar H_i}, d_{\bar H_i}^c,\nu_{\bar H_i}^c);~~i=1,2
\end{equation}
The effective potential in  this theory is\cite{zichichi}   
\begin{eqnarray}
W=\lambda_1 TTh+\lambda_2T\bar F\bar h+\lambda_3\bar F\it l^ch+
\lambda_4 HHh\nonumber\\
~~+\lambda_5 \bar H\bar H\bar h
+\lambda_6 F\bar H\phi +\mu h\bar h\nonumber\\
+\delta_1 HTh +\delta_2 H\bar F\bar h
 +M' H\bar H+ M''\phi \phi
 \end{eqnarray}
where $\phi's$ are the singlet fields. 
Symmetry breaking occurs via VEV formation of $H,\bar H$ fields 
\begin{equation}
<\nu_{H_i}^c>=M_i, <\nu_{\bar H_i}^c>=\bar M_i 
\end{equation}
The Higgs doublet mass matrix arises from  
\begin{equation}
\mu h\bar h\rightarrow \mu H\bar H,~~ \delta_2 H\bar F\bar h
\rightarrow \delta_2 M L \bar H
\end{equation}
and thus has the matrix form 
(with column labelled by $\bar H$
and rows by H and L): 
\begin{equation}
M_{doub}=\pmatrix{\offinterlineskip
 \mu\cr
\delta_2 M\cr}
\end{equation}
The usual scenarios consider the hierarchy: 
$M_i\sim  \bar M_i\sim M >>M'>> \mu$. 
The light Higgs is gotten by setting $\delta_2=0$.
Regarding the Higgs triplet sector, in addition to 
the D and $\bar D$ one also has color triplets from
the 10 and $\bar {10}$ of Higgs. Thus there is mixings between
the D and $\bar D$ terms and the color triplet terms from
the 10 and $\bar 10$ of SU(5). After  symmetry breaking mass
contributions arise as follows:
\begin{eqnarray}
 \mu h\bar h\rightarrow \mu D\bar D\nonumber\\
\delta_1 TFh\rightarrow \delta_1 M d^c D, 
\lambda_4 TTh\rightarrow Md^c_HD\nonumber\\
\lambda_5 \bar H\bar H\bar h\rightarrow \lambda_5 \bar m d^c_{\bar H} \bar
D, M'H\bar H\rightarrow M' d^c_Hd^c_{\bar H}
\end{eqnarray}
Often one sets  $\delta_1=0$. With these assumptions the 
mass matrix in the Higgs triplet sector is
(columns: labelled
by $\bar D$, $d^c_H$; rows: labelled by $D, d^c_{\bar H}$):
\begin{equation}
M_{trip}=\pmatrix{\offinterlineskip
{\mu}&\lambda_4 M \cr
\lambda_5 M   &{M'}\cr}
\end{equation}
Diagonalization gives the Higgs triplets a superheavy mass O(M).
The effective dimension five operator arising from $\mu D\bar D$ is of size
\begin{equation}
\mu/M^2<<1/M
\end{equation} 
 Thus in the flipped model  
 the dimension five operators which mediate proton
decay would be suppressed by a $\mu/M^2$ rather than the usual $1/M$ factor
as in the standard SU(5) model. However, it is possible to upset 
this suppression in some cases. 

\subsection{\bf Proton Decay in SO(10) Models }
SO(10) models have several interesting features. One of these is that
SO(10) allows for the doublet-triplet splitting via VEV alignment.
To achieve this one considers a Higgs sector of the type\cite{dimo} 
\begin{equation}
W_{d-t}=10_145_110_2+ 10_2^2\phi
\end{equation}
where $10_1$ is the Higgs that couples with matter and $\phi$ is 
a singlet. $45_1$ gets a VEV  O($M_G$) in the B-L direction 
giving the mass matrices 

\begin{equation}
M_{doub}=\pmatrix{\offinterlineskip
{0}& 0 \cr
0  &{\phi}\cr}
\end{equation}
\begin{equation}
M_{trip}=\pmatrix{\offinterlineskip
{0}&a_1 \cr
-a_1   &{\phi}\cr}
\end{equation}
The Higgs doublet is  light but the higgs 
triplets develop an effective mass 
\begin{equation}
M_{PD}^{-1}=M_{11}^{-1} =\phi/a_1^2
\end{equation}
In proton decay analyses in SO(10) models one has in addition
to the matrix element $\beta_p$ defined by Eq.(35) also 
the matrix element $\alpha_p$ defined by 
\begin{equation}
\alpha_p U_L^{\gamma}=\epsilon_{abc}\epsilon_{\alpha \beta} 
<0|\bar d_{aL}^{\alpha} \bar u_{bL}^{\beta}u_{cL}^{\gamma}|p>
\end{equation}
where $|\alpha_p|=|\beta_p|$ but the phase of 
$\alpha_p/\beta_p$ remains to be specified.
This phase can affect proton decay rates\cite{lucas}. 

 SO(10) models generally involve large $\tan\beta$. 
 Because of that there are some features specific to  
 these models. Thus, for example, one finds that the
 gluino contributions can be very significant and comparable
to the contributions from the chargino exchange. Further, it is 
found that the contributions of the dimension six LLRR 
operators may be comparable to or even 
dominate  the LLLL contribution in these models\cite{lucas}. 
However, there is a potential 
problem in SO(10) regarding proton stability vs 
 unification of gauge couplings using LEP data. To see the problem 
 one notices that the mass scale necessary to suppress p decay to the 
 current experimental value is\cite{testing}

\begin{equation}
(M^{-1})_{11}> tan\beta (0.57\times 10^{16})~ GeV
\end{equation}
which for tan$\beta\sim 50$ requires 
a GUT mass of 2.5$\times 10^{17}$ for suppression of p decay
to the current experimental limit.
However, a mass scale of this size upsets unification of gauge 
couplings and one  needs large threshold corrections to get 
agreement with experiment\cite{urano,barr1}. 
Fig.2 illustrates the problem of the high scale in SO(10). One
finds from Fig.2 that for the experimental value of 
$\sin^2\theta_W(M_Z)_{\bar{MS}}$ which is $0.23122\pm 0.00002$ 
the value of $\alpha_s$ corresponding to $M_{PD}=2.7\times 10^{17}$
is many standard deviations away from the current experimental
limit of $\alpha_s(M_Z)=0.118\pm 0.003$. Thus as stated large
threshold corrections are necessary to get consistency.
There are a great variety
of SO(10) models, and details of the model affect critically the
nature of supersymmetric signals. Thus in a class of SO(10) models 
discussed recently\cite{bpw} the charged lepton modes of the proton
such as $\it l^+\pi^0$, $\it l^+K^0$ and $\it l^+\eta$ ($\it l =e,\mu$),
can become prominent\cite{bpw}. There are also other approaches 
in the non-minimal extensions where proton decay is correlated 
directly with the ansatz on the Yukawa couplings of the dimension five 
operators\cite{strassler}.

\begin{figure}
\vskip -1.5cm
\psfig{figure=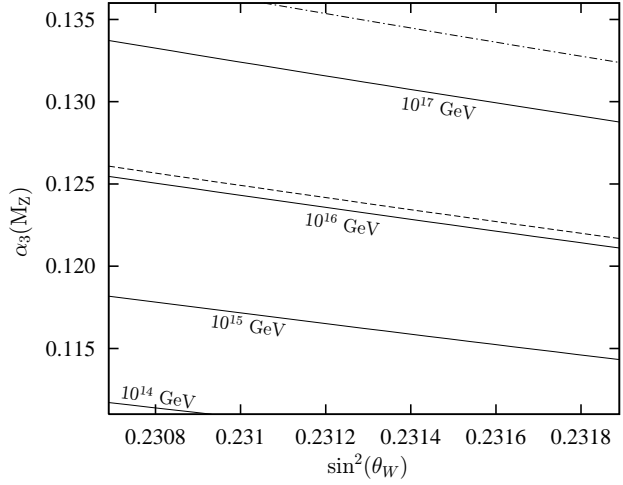,height=2.5in}
\caption{ Plot in the $sin^2\theta_W-\alpha_3(M_Z)$ plane for
various $M_{PD}$ values. The dashed line is for the case of 
the lower bound on $M_{PD}=1.2\times 10^{16}$ GeV in the minimal 
SU(5) model.  The dot-dashed line corresponds to the lower bound on
$M_{PD}=2.7\times 10^{17}$ GeV in the minimal SO(10) model.
 (Taken from ref.[45]).
\label{fig:radis2} }
\end{figure}

\section{Effects of Textures on P Decay}
	Gut models give poor predictions for quark-lepton mass ratios. 
	In SU(5) $m_b/m_{\tau}$ is in good agreement with 
	experiment but $m_s/m_{\mu}$ and $m_d/m_e$ are not. One needs 
	textures, i.e.,    
	
	\begin{equation}
	W_{d}=H_1 l A^Ee^c+H_1d^c A^Dq+ H_2u^c A^Uq
	\end{equation}
	where $A^E, A^D, \& A^U$ are the texture matrices. 
	The simplest choice for these are the 
	the Georgi-Jarlskog choice\cite{georgi} 

	\begin{equation}	
       	A^E= \left(\matrix{0&F&0\cr
                  F&-3E&0\cr
                  0&0&D\cr}\right),
  	 A^D=\left(\matrix{0&Fe^{i\phi}&0\cr
                  Fe^{-i\phi}&E&0\cr
                  0&0&D\cr}\right)                
 	   \end{equation}

	\begin{equation}
	A^U=\left(\matrix{0&C&0\cr
                  C&0&B\cr
                  0&B&A\cr}\right)
	  \end{equation}                
where  A-F  have a hierarchy of the type 
\begin{eqnarray}
A\sim ~O(1),~~B,D\sim ~O(\epsilon)\nonumber\\
C,E\sim ~O(\epsilon^2),~~F\sim O(\epsilon^3),~~\epsilon~<<~1
\end{eqnarray}
\noindent
A possible origin of the parameter $\epsilon$ is from the ratio of
mass scales, e.g., 
\begin{equation}
\epsilon=\frac{M_{GUT}}{M_{str}}
\end{equation}
In the context of supergravity unified models this ratio can arise
from higher dimensional operators.
In the energy domain below the string scale after integration over
the heavy modes of the string one has an effective theory of the type
 \begin{equation}
 W=W_3+W_4+W_5+...
 \end{equation}
where $W_n (n>3)$ are suppressed by the string (Planck) scale  and  
in general contain the adjoints which develop 
VEVs$\sim$ $O(M_{GUT})$. After VEV formation of the  heavy fields 
\begin{equation}
W_n\sim O(\frac{M_{GUT}}{M_{string}})^{n-3}\times ~operators ~in ~W_3
\end{equation}
With the above one can generate mass heirarchy with $\lambda_{yuk}\sim 
O(1)$. One can also compute textures in the Higgs triplet sector

\begin{equation}
	W_t=H_1 l B^E q + H_2 u^c B^U e^c\nonumber\\ 
	+\epsilon_{abc} (H_1 d^c_b B^D u^c_c + H_2^a u^c_b C^U d_c)
\end{equation}
However, the Planck scale expansion is not unique and consequently 
the Higgs triplet texture are not unque, and a dynamical principle 
is needed to achieve uniqueness. One such principle is
the assumption of an exotic sector wherein fields in a minimal
vector like representation couple to both the hidden sector
fields and fields in the physical sector. If the exotic fields
gain superheavy masses their elimination will lead to a spcific
set of Planck scale interactions.  Such an assumption indeed
leads to unique Higgs triplet textures of the form\cite{nathprl}

 \begin{equation}
B^E=\left(\matrix{0&aF &0\cr
  a^*F &{16\over 3}E &0\cr
                  0&0&{2\over 3}D\cr}\right)
   \end{equation}               

\begin{equation}
B^D=\left(\matrix{0&-{8\over 27}F &0\cr
       {-8\over 27}F &-{4\over 3}E &0\cr
                  0&0&-{2\over 3}D\cr}\right)
                  \end{equation}
\begin{equation}
B^U=\left(\matrix{0&{4\over 9}C &0\cr
                  {4\over 9}C &0&-{2\over 3}B\cr
                  0&-{2\over 3}B&A\cr}\right)
     \end{equation}             
    \begin{equation}              
C^U=B^U,~~a=(-{19\over 27}+e^{i\phi})
\end{equation}
 Inclusion of textures gives a moderate
		modification of p  decay branching ratios. 
		The p lifetime for the $\bar \nu K^+$  mode is 
		enhanced by a factor of 
		$\sim (\frac{9}{8}\frac{m_s}{m_{\mu}})^2$.
p decay modes hold important information on GUT physics. Further,
		the textures affect in a differential way the 
		various decay modes which in turn can be used to  
		test theories of textures\cite{nathprl,bb}.

\section{Effects of Dark Matter on Proton Stability}
In this section we discuss the effects of dark matter constraints on 
proton stability. There is now a great deal of convincing evidence for the 
existence of dark matter in the universe and most of this dark matter must be
non-baryonic.  In supersymmetric  theories with R parity invariance the least 
massive supersymmetric particle (LSP) is absolutely stable and hence a candidate 
for dark matter. A priori there are many possible LSP candidates for such 
non-baryonic dark matter in supersymmetry such as the gravitino, 
sneutrino, neutralino, etc. However, detailed analyses show that over most of the
parameter space of the SUGRA  models, it is the 
lightest neutralino ($\chi_1$) which is 
 the LSP and thus a candidate for cold dark matter.
  The LSP neutralino is an admixture of four neutral states, i.e.,  
 \begin{equation}
 \tilde{\chi}_{1}=n_{1}\tilde{W}_{3}+n_{2}\tilde{B}+n_{3}\hat{H}_{1}+n_{4}
\tilde{H}_{2}
\end{equation}  
where $\tilde B$ is the Bino, $\tilde W_3$ is the neutral Wino, and $\tilde H_1, \tilde H_2$ are
 the neutral Higgsinos. In the scaling region when $\mu^2>>M_Z^2$ one finds 
 \begin{eqnarray}
n_1\simeq -\frac{M_Z^2}{2m_{\chi_1\mu}} sin2\theta_W sin\beta,
n_2\sim 1-\frac{1}{2}\frac{M_Z^2}{\mu^2} sin^2\theta_W\nonumber\\
n_3\sim -\frac{M_Z}{\mu} sin\theta_W sin\beta,
n_4\sim \frac{M_Z}{\mu} sin\theta_W cos\beta
\end{eqnarray}
 The above analysis shows that the LSP is mostly a Bino in the
  scaling region which is most of the
 parameter space of the model. In the region when $\mu$ is
 small scaling breaks down, and one can have a large Higgsino component
 for the LSP.  However, recent analyses show that this possibility 
 may be close to being eliminated\cite{falk}.

\subsection{Cosmological Constraints }
The amount of dark matter in the universe puts rather stringent constraints on 
supersymmetric models\cite{lopez2,accurate}.
 These constraints arise from the allowed range of
 $\Omega_{\chi_1^0}h^2$ where 
 $\Omega_{\chi_1^0}=\rho_{\chi_1}^0/\rho_c$. Here
$\rho_{\chi_1}^0$ is the neutralino matter density, and 
 $\rho_c$ is the critical matter density  
\begin{equation}
\rho_c=3H_0^3/8\pi G_N=1.88 h_0^2\times 10^{-29}gm/cm^3
\end{equation}
and $h_0$ is the Hubble parameter $H_0$ in units of 100 km/s.Mpc. 
The  number density  of $\chi_1$'s  obeys the Boltzman equation 
\begin{equation}
\frac{dn}{dt}=-3Hn-<\sigma v>(n^2-n_0^2)
\end{equation}
where $<\sigma v>$ is the thermal average of the neutralino 
annihilation cross-section and v is the relative neutralino 
velocity. 
 At the "freeze-out" temperature $T_f$ the $\chi_1$ 
 decouple from the background, 
 and integration  from $T_f$ to the current temperature 
gives 
\begin{equation}
\Omega_{\tilde\chi_1^0} h^2\cong 2.48\times 10^{-11}{\biggl (
{{T_{\tilde\chi_1^0}}\over {T_{\gamma}}}\biggr )^3} {\biggl ( {T_{\gamma}\over
2.73} \biggr)^3} {N_f^{1/2}\over J ( x_f )}
\end{equation}
where $J~ (x_f) = \int^{x_f}_0 dx ~ \langle~ \sigma \upsilon~ \rangle ~ (x)
GeV^{-2}$,  $N_f$ is the number of massless degrees of freedom at the freezeout, 
$x_f= kT_f/m_{\tilde{\chi}_{1}}$,
 $ T_{\gamma}$ is the current background temperature,
and  $(T_{\tilde\chi_1^0}/T_{\gamma})^3$ is the reheating factor.
 In the anaysis below we shall assume the relic density constraint of 
\begin{equation}
0.1\leq \Omega_{\chi_1^0}h^2\leq 0.4
\end{equation}
 which encompasses a wide range of cosmological models.
 The imposition of the dark matter constraint  reduces the 
 proton lifetime by a very significant amount for values of
 the gluino mass greater than about 500 GeV\cite{limits}. 
 The reduction factors
 as a function of the gluino mass are listed in Table 2 (taken from
 ref.\cite{limits}). One finds
 that the reduction factors lie in the range 10-30. 
 Taking into account corrections due to Yukawa textures, 
 and other uncertainties in the calculations, SuperK should 
 be able to test the region $m_{\tilde g}>500$ GeV when it 
 achieves a sensitivity of $1\times 10^{33}$. 
 These results are exhibited in Fig.3 for
 the minimal supergravity model.
 
\begin{figure}
\vskip -1.5cm
\psfig{figure=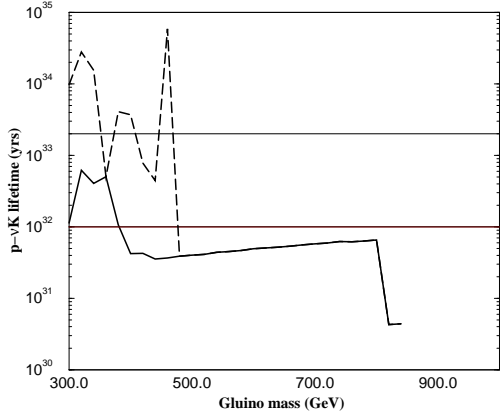,height=2.5in}
\caption{Plots of the maximum $\tau(p\rightarrow \bar \nu K)$ lifetime
	in the minimal SU(5) 
	supergravity model with universal soft breaking as a function
	of the gluino mass with the inclusion of relic density constraint
	on neutralino dark matter of $0.1<\Omega_{\chi_1^0} h^2<0.4$. 
	The analysis is for the naturalness limits on $m_0$ of 
	1 TeV (solid), and 5 TeV (dashed). The lower  
	horizontal solid line is the  Kamionkande limit
	 and the upper horizontal solid line is the lower limit 
	expected from Super K. (Taken from ref.[35]).
\label{fig:radis3} }
\end{figure}

 \subsection{Effects of Proton Lifetime Constraint on Event Rates}
 The imposition of proton lifetime constraint has an important
 effect on the event rates  in dark matter analysis\cite{events}.
 One  finds that in addition to the drastic reduction in the 
 allowed range of the gluino mass, there is also a very significant 
 reduction in the magnitude of the maximum event rate curves.
 In Fig.4 we exhibit the maximum event rate as a function of the 
 neutralino mass. We find that  the reduction of the maximum 
 event rate curve can be a factor of 10 to 100. This reduction
 arises from the fact that the proton lifetime constraint 
 eliminates certain parts of the parameter space which give 
 rise to large event rates.

\begin{figure}
\vskip -1.5cm
\psfig{figure=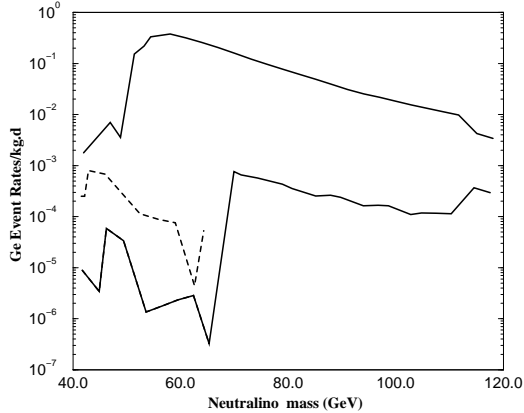,height=2.5in}
\caption{Plots of the 
	the maximum and the minimum of event rates for the scattering 
	of  neutralinos off germanium target as a function of the 
	neutralino mass. The relic density constraint
	 $0.1<\Omega_{\chi_1^0} h^2<0.4$
	is imposed. The solid curves are  for the case of  $m_0\leq 1$ TeV
	and without p decay constraint. The dashed curves
	are for the case with the same relic density constraint,
	but with $m_0\leq 5$ TeV, and
	under the proton lifetime constraint 
	$\tau (p\rightarrow \bar\nu K)>1\times 10^{32}$ yr. The minimum 
	dashed curve coincides with the lower solid curve. [taken from
	Ref.[35]).
\label{fig:radis4} } 
\end{figure}

{\bf 
\begin{center} \begin{tabular}{|c|c|}
\multicolumn{2}{c}{Table~1: Reduction of $\tau(p\rightarrow \bar \nu K)_{max}$ 
from dark matter constraint\cite{limits}.
   } \\
\hline
gluino mass (GeV)  & reduction factor when $0.1<\Omega h^2<0.4$  \\
\hline
  500 & 29.5   \\
\hline
 550 & 23.2  \\
\hline
 600 & 18.6  \\
\hline
 650 & 15.3   \\
\hline
 700  & 12.9   \\
\hline
 750 & 11.3   \\
\hline
 800 & 9.8  \\
\hline
\end{tabular} 
\end{center}
  }

\section{Proton Stability in Gauge Mediated Breaking of Supersymmetry}
In the simplest model of gauge mediated supersymmetry 
breaking \cite{giudice}(GMSB), one has that 
 SUSY breaking arises from VEV formation of a chiral superfield 
 $\hat S$ which couples to messenger chiral 
 superfield $(\phi_i, \bar\phi_i)$ in vector like representations 
  of the MSSM group with a superpotential of the form 
 \begin{equation}
W=\sum C_iS\phi_i\bar\phi_i
\end{equation}
 where in SU(5) one has  $(\phi, \bar\phi)=(5,\bar 5)$, or $(10,\bar 10)$.
 The information on the breaking of supersymmetry is communicated by 
 gauge interactions to the physical sector.
 Masses for the gauginos ($\tilde M_a$) arise at the one loop level via
 gauge interactions, i.e., 
 \begin{equation}
 \tilde M_a=\frac{\alpha_a}{4\pi}\Lambda,~~\Lambda=\frac{<F>}{<S>}\sim 10^2 TeV
 \end{equation}
 where $\sqrt F$ is the SUSY breaking scale, $\Lambda$ is the 
 cutoff and typically $\sqrt F\sim \Lambda$. 
 Masses for the  scalars arise at the two loop level, i.e,
 \begin{equation}
 \tilde m^2_i=\sum_a 2C_{iF}^a(\frac{\alpha_a}{4\pi})^2\Lambda^2
  \end{equation}
  where $C_{iF}^a$ are the Casimir co-efficients for the field i. 
 
 There are many extensions of this simplest GMSB version including 
 models with the messenger fields in incomplete 
 multiplets\cite{martin}.
  However, in  these  scenarios it is difficult to 
  generate  $\mu$ and $B\mu$ terms (where $B\mu$ is the value of
  $B_0\mu_0$ at the electro-weak scale), 
  and  one needs non-gauge interactions for their generation.
   Further, because of the low value of supersymmetry breaking scale
  $\sqrt F$ in these theories, i.e., $\sqrt F\sim 10^2$ TeV, the
   gravitino is the LSP with a mass  
  $m_{3/2}< 1$ Kev\cite{primack}. 
  Thus  the gravitino cannot be a CDM in these models.
  
  There are significant constraints that arise in these models in 
  a GUT framework. First it is noted that the
  unification of the gauge couplings using the LEP data along with
  the $b-\tau$ unification puts severe constraints on the  
  models\cite{carone}. This type of constraint, however, can be 
  softened by inclusion of the Planck scale corrections as in
  Refs.\cite{das,ring}.
  However, it is further found that the limits on  $tan\beta$ from
  proton stability and the limits on it from radiative breaking of the
  electro-weak symmetry under the constraint that the bilinear and
  the trilinear soft couplings vanish at the messenger scale 
  eliminate all GUT models considered in ref.\cite{hamidian} except 
  for one or two isolated cases.  

\section{Planck Effects and Proton  Decay}
In the R-G analysis of the gauge coupling constants $\alpha_i$ 
there is an overlap of GUT threshold effects and of the 
Planck scale effects. For example, for SU(5) and for $Q\sim M_G$
one has\cite{das,ring} 
\begin{equation}
\alpha_i^{-1}(Q)=\alpha_G^{-1}+C_{ia}ln(\frac{M_a}{Q})+
\frac{cM}{2M_P}\alpha_G^{-1}n_i
\end{equation}
where $M_a$ are the superheavy GUT thresholds, and 
$n_i=(-3,-1,2)$ for the SU(2), U(1) and $SU(3)_C$ sectors.
It is easily seen that the  Planck effects characterized by 
$n_i$  can be absorbed in the GUT thresholds by 
 a rescaling. Thus for the minimal SU(5)   
 model rescaling gives  
 \begin{equation}
 \alpha_i^{-1}(Q)=\alpha_G^{eff-1}+C_{ia}ln(\frac{M_a^{eff}}{Q})
 \end{equation}
where
\begin{equation}
 M_a^{eff}=M_ae^{k_aC_P}; k_a=(-\frac{3}{5}(\Sigma),\frac{3}{10}(V),
 5(H_3))
 \end{equation}
 and  where
 \begin{equation}
\alpha_G^{eff-1}=\alpha_G^{-1}-\frac{15}{2\pi}C_P,~~C_P=\frac{\pi cM}
{\alpha_GM_P}
\end{equation}
While the RG analysis involves the effective parameters
$M_V^{eff}$ and $M_{H3}^{eff}$,
the proton decay lifeitme is determined by $M_V$ and by $M_{H_3}$.
Thus the RG analysis along with 
proton lifetime measurements can allow one to measure the size of
the Planck correction, i.e., the value of c. 
For the $p\rightarrow \bar \nu K^+$ one finds\cite{ring} 
\begin{equation}
p\rightarrow \bar \nu K^+, ~~c=\frac{\alpha_G}{10}\frac{M_P}{\pi M_V}
ln\frac{M_{H_3}^{eff}}{M_{H_3}}
\end{equation}
and for the $p\rightarrow e^+\pi^0$ mode one finds\cite{das} 
\begin{equation}
p\rightarrow e^+\pi^0,~~c=\frac{100}{3}\sqrt{\frac{2}{3}}
\alpha_G^{3/2}\frac{M_P}{M_V}ln\frac{M_V}{M_V^{eff}}
\end{equation}

\section{Exotic p Decay Modes}
p decay modes discussed so far are all  of the type where a proton decays 
into an anti-lepton and a meson, i.e., 
\begin{eqnarray}
p\rightarrow e^+\pi^0\nonumber\\
 p\rightarrow \bar \nu K^+(\pi^+)\nonumber\\
 p\rightarrow \mu^+K^0
 \end{eqnarray}
 These decay modes arise in the minimal SU(5) and SO(10) models
 where the interactions obey the $B-L$ conservation. However, it 
 is possible to include interactions where $B-L$ coservation 
 is violated. Thus, for example, one may consider an interaction 
 of the type\cite{zee}, 
  \begin{equation}  
  5_M5_M\bar 10_H
  \end{equation}  
  which  can generate $\Delta (B-L)=2$ transitions. Such interactions
  will allow  proton decay modes with 
   a lepton and mesons such as  
   \begin{eqnarray}
  d+d+s\rightarrow \mu^-\nonumber\\
  n\rightarrow \mu^-K^+\nonumber\\
  p\rightarrow \mu^-\pi^+K^+,..\nonumber\\
\end{eqnarray}
    Thus p decay modes distinguish among the varieties of GUT 
    interactions and can provide important insights into 
    GUT physics.
  
\section{Connection with String/M Theory}
Proton stability is a very strong constraint on string model 
building. Most string models contain interactions which violate
R parity and can generate rapid proton decay. However,
even with R parity invariance models with string GUTs will show
generically the same typical B and L violating interactions that
one has in ordinary GUT models. Recently, there has been an 
exhaustive study of string theories which perturbatively 
 allow grand unification.
One finds that  indeed it is possible to find string models
with interesting unified gauge groups such as SU(5), SO(10),
E(6) etc\cite{lewellen}. However, simultaneous satisfaction of other 
desirable properties such as $N=1$ space time supersymmetry,
three chiral families, and massless ajoints needed to
break the gauge symmetry is not  so easy, and one needs to
go to models with  higher Kac-Moody levels for their satisfaction.
Recently, there has been a great deal of work on models of this
type and explicit models at Kac-moody level three with the
above properties have been constructed\cite{kaku}. Unfortunately, there
are several problems of a phenomenological level still to
be overcome before such models can become viable.
String Guts is one of the many ways in which one can reconcile 
LEP data and the unification of gauge couplings within string theory. 
Without string GUTS one will have the Standard Model gauge group 
emerging directly at the string scale, and to reconcile the LEP
data on the gauge coupling constants with this high scale, one
needs some extra effects, such as string thresholds and extra vector
like representations at an intermediate state below the 
string scale\cite{dienes}.

Horava and Witten\cite{horava,witten} have suggested a new possibility
for the unification of the gauge couplings within the 
framework of string/M theory. It is within the framework of the
conjecture that the
strongly coupled limit of the $E_8\times E_8$ heterotic string
is M theory on $R_{10}\times S_1/Z_2$ with gravity propagating
in the bulk and the gauge fields living on the 
hyperplanes\cite{horava,ellis}.
If M theory compactifies  on an $M_4\times CY\times S_1/Z_2$
the unification of gauge coupling constants
can arise at the GUT scale with the MSSM spectrum with the appropriate
identification of the Calabi-Yau compatification radius with the 
inverse of the GUT scale\cite{witten}.  One has
in addition unification of gauge coupling constants with gravity
arising at this scale because of an extra running of the
gravitational coupling due to the opening up of the fifth dimension
at a scale which is an order of magnitude below the GUT scale.
This picture is very close to the supergravity GUT picture as far
as the particle sector of the theory is concerned.

\section{Conclusions/Prospects}
LEP data appears to support ideas of both grand unification and of
supersymmetry. Thus SUSY/SUGRA GUT may be an important way station to the 
Planck scale where unification of all interactions occurs.
In SUSY/SUGRA GUTS one needs R parity invariance 
to eliminate B and L violating
dimension 4 operators which lead to rapid proton decay. 
  B and L violating dimension 5 operators of GUT models allow  
one to probe via proton decay a majority of the parameter space 
of the minimal SUGRA model within the naturalness constraint of 
$m_0\leq 1$ TeV and $m_{\tilde g}\leq 1$ TeV if the Super-K and 
ICARUS experiments 
can reach the expected sensitivity of  $2\times 10^{33}$ y and
$10^{34}$ y respectively for the  $\bar\nu K^+$ mode.
With the inclusion of dark matter constraints one finds that 
 if the proton decay is not observed the gluino mass must lie  
below 500 GeV within any reasonable naturalness constraint.
The simultaneous p stability and dark matter constraints  will be 
tested in the near future in p decay experiments as well by 
 experiments at the Tevatron, LEP2 and the LHC.

\section*{Acknowledgments}
This work was supported in part by NSF grant number PHY-96020274
and PHY-9722090.

\end{document}